\documentclass {article}
\usepackage{graphicx} 
\usepackage{amsmath}
\usepackage{amsfonts}
\usepackage{amsmath}
\usepackage{amssymb}
\usepackage{graphicx,bm,units,yfonts}
\usepackage{subfigure}
\usepackage[table]{xcolor}
\usepackage{textcomp}
\usepackage{textcomp}
\usepackage{soul}
\usepackage{hyperref}
\usepackage{svg}
\usepackage{float}
\usepackage[normalem]{ulem}
\usepackage{authblk}

\usepackage{setspace}

\usepackage[margin=1in]{geometry}

\newcommand{\Ll}{{\mathcal L}}

\title{Topological Acoustic Metamaterials with Synthetic Magnetism: Observation of Hofstadter Spectrum and Topological Edge States }

\author{
Elena Rybalova \textsuperscript{1, *}
Jackson Saunders \textsuperscript{1}, 
Curtis Rasmussen \textsuperscript{2}, 
Nicholas H. Patino \textsuperscript{2},
Emil Prodan\textsuperscript{3}, 
Massimo Ruzzene \textsuperscript{2},
Camelia Prodan\textsuperscript{1, *}
}

\date{}  

\begin{document}

\maketitle

\begin{center}

\textsuperscript{1} Department of Physics and Engineering Physics, Fordham University, New York NY 10458, USA. \\

\textsuperscript{2} Department of Mechanical Engineering, University of Colorado Boulder, Boulder, Colorado 80309, USA.\\

\textsuperscript{3} Department of Physics, Yeshiva University, New York, NY 10033, USA. \\

\bigskip
\textsuperscript{*}Corresponding authors: erybalova@fordham.edu, cprodan@fordham.edu
\end{center}

\doublespacing

\section{Abstract}

Architected materials are increasingly used to implement topological dynamics, yet realizing the gauge fields and symmetries that protect the topology with purely passive elements remains the central practical challenge. Here we turn a previously proposed theoretical strategy into an integrated platform that accepts a theoretical model as input and returns a passive acoustic metamaterial with identical dynamics. We demonstrate it experimentally on the Hofstadter model at $\theta = 2\pi/3$, fabricating a spin-Chern insulator whose two sectors carry Chern numbers $C_s = \pm 1$, and observe the predicted helical edge states.

\section{Introduction}

The classification of topological phases of matter  sorts gapped systems by the symmetries they respect and the integer or
$\mathbb{Z}_2$ invariants they carry~\cite{haldane2017nobel,stanescu2024intro,ryu2010,chiu2013}.
The bulk--boundary correspondence, a central feature of such systems, says that a nontrivial bulk invariant forces protected states onto the system's
edges, states that carry excitations without backscattering and that
are immune to any disorder which does not close the bulk
gap~\cite{prodanSB2016,hatsugai1993}. Because this behavior is dictated
by the structure of an effective Hamiltonian rather than by any
intrinsically quantum ingredient, it can be realized in classical
wave systems whose dynamical matrices reproduce the target Hamiltonian.
Photonic~\cite{krishnamoorthy2023,raghu2008,haldaneRaghu2008},
acoustic~\cite{maSheng2016,yang2015,chenYF2023,xue2022}, and
mechanical~\cite{chenAdvMech2021,huang2021} metamaterials have
consequently become versatile, room-temperature platforms for
topological band physics~\cite{maXiaoChan2019,ozawa2019}, with
coupled-resonator architectures offering an especially direct dictionary
between lattice geometry and Hamiltonian matrix
elements~\cite{chenSchrodinger2025,abood2025,lin2023,wuY2023}.

Within the tenfold classification, Class~A---distinguished by the
complete absence of time-reversal, particle-hole, and chiral
symmetry---is the symmetry class of the integer quantum Hall effect, and
in two dimensions it hosts phases indexed by an integer Chern number
whose nonzero value guarantees chiral edge
transport~\cite{ryu2010,chiu2013,thouless1982}. Its canonical lattice
realization is the Hofstadter model, a charged particle hopping on a
square lattice threaded by uniform magnetic flux, whose spectrum traces
the celebrated fractal ``butterfly'' as the flux is
varied~\cite{hofstadter1976}. Realizing Class~A topology in a
metamaterial is uniquely demanding: the flux enters the Hamiltonian as
\emph{complex} hopping phases, and it is precisely these complex phases
that encode the broken time-reversal symmetry the class requires. A
passive acoustic crystal possesses no mechanism to generate them---in
the high-contrast, coupled-mode regime its dynamical matrix is real and
symmetric~\cite{lee2022}, so the Peierls phases of the Hofstadter model
cannot be written down directly. The acoustic and mechanical Chern
insulators demonstrated to date have therefore broken time-reversal
symmetry by brute force, through active or externally biased elements:
angular-momentum--biased ring resonators and circulating
background airflow~\cite{khanikaev2015, ding2019}, spontaneous chiral flow in
active liquids~\cite{souslov2017}, gyroscopic
coupling~\cite{wang2015,nash2015}, and, more recently, programmable
synthetic magnetism in optomechanical resonator
networks~\cite{slim2025}. These schemes are powerful, but each is
experimentally intricate and tailored to a single phase rather than to
the classification table as a whole.

A conceptually distinct route removes the need for active elements
entirely. Barlas and Prodan showed that \emph{any} complex Hamiltonian in
the topological classification table can be implemented with passive,
real-valued couplings by \emph{realifying} it: doubling the internal
degrees of freedom at each site and replacing the imaginary unit with
the $2\times2$ rotation generator
$\left(\begin{smallmatrix}0&1\\-1&0\end{smallmatrix}\right)$~\cite{barlasProdan2018}.
The resulting real Hamiltonian preserves the spectrum, the symmetry
class, and the full boundary spectrum of the original complex model, at
the sole cost of a two-component internal structure per
site~\cite{prodanSB2016,barlasProdan2018}. Applied to the Hofstadter
model, realification yields a bilayer lattice in which the synthetic
flux is carried entirely by the magnitudes and signs of real hoppings.
An apparently quantum, time-reversal--breaking problem thereby becomes a
purely geometric design task---provided one can build a resonator that
carries two coupled internal modes and a coupling scheme that sets each
hopping sign at will. This passive-realification strategy has recently
been validated experimentally for the particle-hole-symmetric D~class,
where real-valued couplings in an enriched unit cell reproduce a pair of
sectors with opposite Chern numbers~\cite{wuSQ2024,chenProdan2024}.

A related passive route arises from projective-symmetry theory, in which sign-alternating real couplings act as a $\mathbb{Z}_2$ gauge field that modifies the algebra of crystalline symmetries and time reversal \cite{Zhao2021,Shao2023}. It has enabled acoustic projectively enriched, Möbius, spinful, and mirror-Chern phases \cite{Xue2022PRL,Li2022,Meng2023,Li2023}. In particular, a bilayer of Hofstadter models with opposite fluxes was recently realized in an acoustic system at $\phi = 2\pi/5$ \cite{Xiang2026}, a zero-mode scheme for arbitrary $O(N)$ gauge links with positive couplings only realized an $SO(2)$ Hofstadter model at $\alpha = \pi/2$ \cite{Dai2026}, and layer degrees of freedom have been used to emulate $SU(2)$ link matrices with real couplings, yielding acoustic Abelian and non-Abelian Hofstadter insulators with full-wave butterflies and measured helical boundary states~\cite{Wang2026}.

In the acoustic setting, the Hofstadter spectrum itself was previously
accessed by our group through a complementary, dimensionally reduced
route: a reconfigurable one-dimensional quasiperiodic crystal whose
almost-Mathieu (Harper--Aubry--Andr\'e) spectrum maps out the full
butterfly and its topologically pumped boundary
states~\cite{ni2019,apigo2019}. 
Although developed from different starting points (projective symmetry, sublattice zero modes, layer pseudospin) the two-dimensional realizations of Refs.~\cite{Xiang2026,Dai2026,Wang2026} share one structure: real-coupled copies whose complex combinations decouple into conjugate sectors, which is precisely the realification of Ref.~\cite{barlasProdan2018}. They differ in its cost and its ingredients: two cavities per site but dedicated tube geometries for each sign \cite{Xiang2026}, six-cavity artificial atoms whose butterfly is embedded between additional high-energy bands \cite{Dai2026}, or four cavity layers per site encoding a spin-1/2 link \cite{Wang2026}. What has not yet been shown is a direct implementation of the realified $U(1)$ Hofstadter model that combines two degrees of freedom per site with a single sign mechanism for all hoppings, resolves the butterfly across the full flux domain on a finite open lattice together with its edge states, and tests explicitly the sector structure on which their protection rests.

Here we realize precisely such a platform. We introduce a double-bridge
H-shaped acoustic resonator whose anti-symmetric  operating mode lets tube connectivity fix the sign of each
hopping---like-cavity links yielding a positive coupling, cross-cavity
links a negative one. Two such resonators per lattice site, one in each layer, furnish the two-component internal space demanded by the realified Hamiltonian. The coupler length is chosen so that hybridization splits the resonance symmetrically, leaving onsite frequencies unchanged as hopping magnitudes vary; because magnitude and sign are thus set independently, the full geometry for any admissible flux is generated algorithmically from the tight-binding Hamiltonian. 
Using COMSOL Multiphysics~\cite{comsol}, cross-validated against analytical
diagonalization of the tight-binding model, we assemble bilayer lattices
that implement the realified Hofstadter Hamiltonian at arbitrary
synthetic flux. Sweeping the gauge parameter $\theta$ across its
fundamental domain, we recover the complete Hofstadter
butterfly---its self-similar gap structure and $E(\theta)=E(\pi-\theta)$
mirror symmetry---in quantitative agreement between simulation and
theory, echoing the fractal spectra recently resolved spectroscopically
in moir\'e superlattices~\cite{dean2013,nuckolls2025} and in
one-dimensional phononic structures~\cite{daviesMorini2026}, and the
historic quantum Hall measurements in lateral
superlattices~\cite{albrecht2001}.

As an example, fixing $\theta = 2\pi/3$, we show that the eigenspectrum of a $9\times9$ bilayer lattice separates into bulk bands and spectral gaps, and that the gap bands host modes sharply localized at the lattice boundary. We further show that these edge states survive the removal of an entire magnetic unit cell from both layers, which cuts a notch into the boundary. In this case, the edge modes follow the deformed boundary without scattering into the bulk, because the same sites are removed in both layers, the perturbation preserves the sector structure on which their topological protection rests~\cite{prodanSB2016,hatsugai1993}. Moreover, the two decoupled sectors carry opposite band Chern numbers, $\pm(+1,-2,+1)$, so that the total Chern number vanishes while a spin Chern number $C_s = \pm1$ protects a helical pair of edge channels in each gap. Finally, we validate these results experimentally at $\theta = 2\pi/3$ using 3D-printed samples, with and without the boundary defect.

Beyond this specific demonstration, our results establish the
double-bridge H-resonator as a passive, reconfigurable building block
for synthetic gauge fields in sound, and realification as a practical
route to the broader topological classification table without active
components, temporal modulation, or moving media.

\section{Imaginary to Real: Doubling Classical Degrees of Freedom to Emulate the Hofstadter Hamiltonian}

The Hofstadter Hamiltonian describes the motion of a charged particle hopping on a 2D square lattice threaded by a uniform perpendicular magnetic field \cite{hofstadter1976}. It is written as
\begin{equation}\label{eq:Hofstadter Hamiltonian}
H = \sum_{\langle \bm x, \bm y\rangle } e^{\imath \theta \, \bm x \wedge \bm y} |\bm x\rangle \langle \bm y|, \quad \bm x \wedge \bm y := x_1 y_2 - x_2 y_1,
\end{equation}
where the sum runs over nearest-neighbor pairs $\bm x=(x_1,x_2)$ and $\bm y=(y_1,y_2)$ on a square lattice $\Ll$ and the parameter $\theta$ encodes the magnetic flux per unit cell. The integer quantum Hall effect---the first experimentally observed topological phenomenon in condensed matter---is canonically modeled by this Hamiltonian.

Passive acoustic crystals operating in the high-contrast regime, where coupled-mode theory applies \cite{lee2022}, possess dispersion relations governed by dynamical matrices of the general form
\begin{equation}\label{eq:Hofstadter HamiltonianA}
    H_{\rm acoustic} = \sum_{\langle \bm x, \bm y\rangle} w_{\bm x,\bm y} \otimes |\bm x \rangle \langle \bm y |,
\end{equation}
in which $w_{\bm x,\bm y}$ are coupling matrices whose dimension equals the number of resonant modes per cavity within the frequency window of interest. Because no active elements are present, the entries of $w$ are necessarily real, and the complex phase factors appearing in Eq.~\eqref{eq:Hofstadter Hamiltonian} cannot be directly reproduced.

This obstacle is overcome by the realification technique introduced by Barlas and Prodan \cite{barlasProdan2018}, which provides a systematic procedure for implementing any model from the topological classification table with passive metamaterials. The prescription amounts to the substitutions
\begin{equation}
1 \mapsto \begin{pmatrix}
    1 & 0 \\ 0 & 1
\end{pmatrix} \ \ {\rm and} \ \ \imath : = \sqrt{-1} \mapsto \begin{pmatrix}
    0 & 1 \\ -1 & 0
\end{pmatrix}.
\end{equation}
These replacements preserve all algebraic relations---in particular the symmetry structure and the energy spectrum (with or without boundaries)---at the cost of doubling the number of internal degrees of freedom per site. Applying this map to the Hofstadter Hamiltonian \eqref{eq:Hofstadter Hamiltonian} yields the real-valued bilayer Hamiltonian
\begin{equation}
H_R = \sum_{\langle \bm x, \bm y\rangle } \begin{pmatrix}\cos( \theta \, \bm x \wedge \bm y) & \sin( \theta \, \bm x \wedge \bm y) \\
-\sin(\theta \, \bm x \wedge \bm y) & \cos( \theta \, \bm x \wedge \bm y)\end{pmatrix} \otimes |\bm x\rangle \langle \bm y|.
\label{eq:quantum Hamiltonian}
\end{equation}
This Hamiltonian commutes with the onsite operator {\small $\begin{pmatrix} 0 & 1 \\ -1 & 0\end{pmatrix}\otimes 1$}, which generates a $\mathbb{Z}_2$ symmetry partitioning $H_R$ into two decoupled sectors. If $\Pi_\pm$ denote the projections onto these sectors, then $\Pi_+ H_R \Pi_+$ is unitarily equivalent to the original Hofstadter Hamiltonian \eqref{eq:Hofstadter Hamiltonian}, while $\Pi_- H_R \Pi_-$ corresponds to the same model with the sign of $\theta$ reversed. The two sectors therefore carry opposite Chern numbers, but their dynamics remain fully independent. In practice, this means that Hamiltonian~\eqref{eq:quantum Hamiltonian} encodes two copies of Hofstadter physics with opposite synthetic magnetic fields, each producing the same spectrum and the same topological edge modes as the original complex Hamiltonian. The bilayer structure of $H_R$ maps naturally onto a physical resonator possessing two coupled cavities---exactly the geometry of the double-bridge H-resonator employed in this work.

\section{Acoustic Metastructure Design}

\begin{figure}[b!]
    \centering
    \def\svgwidth{\textwidth}
\begingroup%
  \makeatletter%
  \providecommand\color[2][]{%
    \errmessage{(Inkscape) Color is used for the text in Inkscape, but the package 'color.sty' is not loaded}%
    \renewcommand\color[2][]{}%
  }%
  \providecommand\transparent[1]{%
    \errmessage{(Inkscape) Transparency is used (non-zero) for the text in Inkscape, but the package 'transparent.sty' is not loaded}%
    \renewcommand\transparent[1]{}%
  }%
  \providecommand\rotatebox[2]{#2}%
  \newcommand*\fsize{\dimexpr\f@size pt\relax}%
  \newcommand*\lineheight[1]{\fontsize{\fsize}{#1\fsize}\selectfont}%
  \ifx\svgwidth\undefined%
    \setlength{\unitlength}{699.13259431bp}%
    \ifx\svgscale\undefined%
      \relax%
    \else%
      \setlength{\unitlength}{\unitlength * \real{\svgscale}}%
    \fi%
  \else%
    \setlength{\unitlength}{\svgwidth}%
  \fi%
  \global\let\svgwidth\undefined%
  \global\let\svgscale\undefined%
  \makeatother%
  \begin{picture}(1,0.62267017)%
    \lineheight{1}%
    \setlength\tabcolsep{0pt}%
    \put(0,0){\includegraphics[width=\unitlength,page=1]{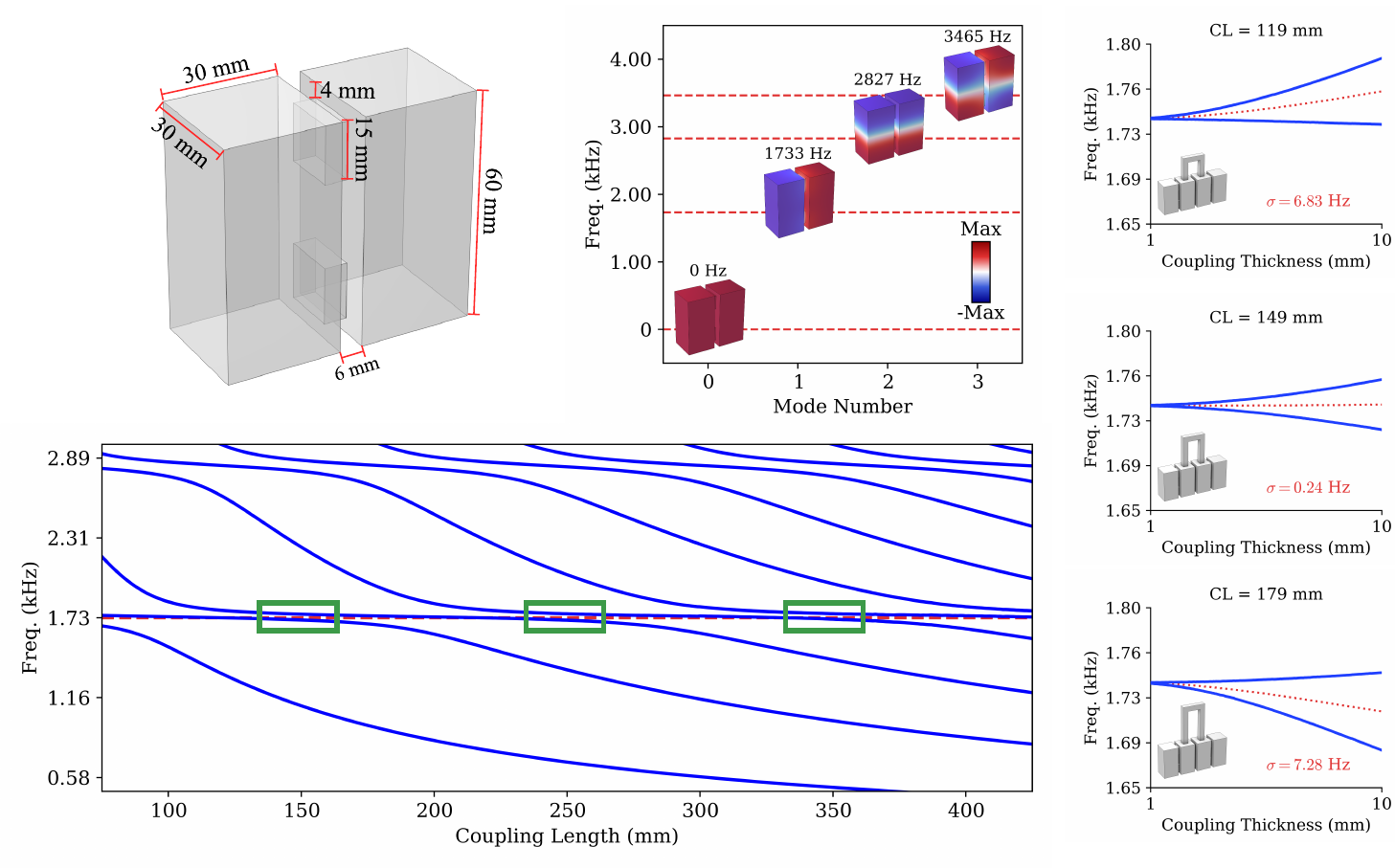}}%
    \put(0.08608609,0.31794683){\makebox(0,0)[lt]{\lineheight{1.25}\smash{\begin{tabular}[t]{l}c)\end{tabular}}}}%
    \put(0.7630146,0.60876622){\makebox(0,0)[lt]{\lineheight{1.25}\smash{\begin{tabular}[t]{l}d)\end{tabular}}}}%
    \put(0.7629364,0.40731833){\makebox(0,0)[lt]{\lineheight{1.25}\smash{\begin{tabular}[t]{l}e)\end{tabular}}}}%
    \put(0.76290701,0.20331566){\makebox(0,0)[lt]{\lineheight{1.25}\smash{\begin{tabular}[t]{l}f)\end{tabular}}}}%
    \put(0.08606514,0.60876622){\makebox(0,0)[lt]{\lineheight{1.25}\smash{\begin{tabular}[t]{l}a)\end{tabular}}}}%
    \put(0.40883931,0.60876622){\makebox(0,0)[lt]{\lineheight{1.25}\smash{\begin{tabular}[t]{l}b)\end{tabular}}}}%
    \put(0,0){\includegraphics[width=\unitlength,page=2]{fig_1.pdf}}%
  \end{picture}%
\endgroup%

\caption{
\textbf{H-resonator and dimer tuning.}
\textbf{a)} Double bridge H-resonator schematic consisting of two chambers connected by two bridges. Crucial parameters are labeled.
\textbf{b)} First 4 modes of the H-resonator (red dashed lines) and the corresponding pressure profiles. The heat map legend shows the minimum and maximum relative acoustic pressure.
\textbf{c)} Eigenfrequencies of a dimer of two coupled H-resonators as a function of coupling length (CL) at a fixed coupling thickness of 10\,mm. The dimer geometry with the CL indicated is shown in the inset. The red dashed line marks the 1733\,Hz operating mode of an isolated H-resonator. Green rectangles highlight the CL intervals in which the splitting of this mode is most symmetric.
\textbf{d)--f)} Eigenfrequency as a function of coupling thickness (ranging from 1\,mm to 10\,mm) for three CLs: 119\,mm, 149\,mm, and 179\,mm. The geometric setup associated with each CL is shown inlaid in the top left corner. The red dotted line represents the average of the upper and lower eigenfrequencies at each coupling thickness. $\sigma$ is the standard deviation of these averaged values across all coupling thicknesses, serving as a measure of symmetry between the upper and lower frequency bands.
}
    \label{fig:1}
\end{figure}

The fundamental building block of the metamaterial is a double-bridge H-shaped resonator, depicted in Figure~\ref{fig:1}(a). It consists of two square cavities joined by a pair of bridging channels, a geometry that naturally furnishes the two internal degrees of freedom required by the $2\times 2$ matrix structure of the realified Hamiltonian~\eqref{eq:quantum Hamiltonian}. Eigenfrequency simulations performed with the acoustics module of COMSOL Multiphysics \cite{comsol} reveal the first four resonant modes of an isolated H-resonator (Figure~\ref{fig:1}(b)). The corresponding pressure profiles confirm well-separated eigenstates, with the lowest nontrivial mode exhibiting anti-symmetric pressure distribution across both cavities. The clear spectral separation between the fundamental and higher-order modes ensures that spurious inter-mode coupling is negligible, preserving the validity of the single-mode tight-binding description.

To assemble the resonators into a lattice, adjacent H-resonators are linked by coupling tubes whose geometric parameters---length and cross-sectional thickness---govern the inter-site hopping strength. The sign of the effective coupling (positive or negative) is controlled by which cavity each tube connects to: coupling to the same cavity in two neighboring resonators yields a positive hopping amplitude, while connecting opposite cavities introduces a sign reversal. This mechanism enables the independent tuning of both the magnitude and the sign of each hopping element in the Hamiltonian, as required by Eq.~\eqref{eq:quantum Hamiltonian}. A detailed description of this positive and negative coupling scheme in bilayer systems can be found in prior work (see, for example, Figure~2 in Ref.~\cite{wuSQ2024}).

Figure~\ref{fig:1}(c) presents a parametric sweep of the dimer eigenfrequencies as a function of coupling length at a fixed coupling thickness of 10\,mm. When two H-resonators are coupled, their degenerate operating modes near 1733\,Hz hybridize into a pair of eigenstates. In addition, the coupling tube supports its own acoustic resonances, whose frequencies decrease with increasing CL. Where such a tube resonance approaches the operating frequency, it repels the dimer modes and shifts the pair asymmetrically. Between these regions, the two dimer modes split nearly symmetrically; these CL intervals recur along the sweep and are highlighted by green rectangles in Figure~\ref{fig:1}(c). Only coupling lengths within these windows leave the onsite frequencies insensitive to the hopping magnitude and are therefore suitable for lattice construction.

To quantify this symmetry and select the coupling length, Figure~\ref{fig:1}(d)--(f) show the dimer eigenfrequencies as a function of coupling thickness (1--10\,mm) for three coupling lengths: 119\,mm, 149\,mm, and 179\,mm. 
The middle value lies inside the first symmetric window of Figure~\ref{fig:1}(c), while the other two lie on either side of it. In all three cases the frequency separation of the pair grows with increasing coupling thickness.
For an ideal lattice implementation, this splitting must be symmetric about a fixed midpoint frequency, so that the resulting band structure accurately reflects the intended Hamiltonian.
To quantify this symmetry, we compute the average of the upper and lower eigenfrequencies at each coupling thickness (red dotted line in Figure~\ref{fig:1}(d)--(f)) and evaluate its standard deviation $\sigma$ across all thicknesses. A small $\sigma$ indicates that the midpoint frequency remains nearly constant regardless of coupling strength, corresponding to a well-balanced energy distribution between the two resonators. Among the three coupling lengths tested, the 119\,mm coupler produces a slight upward bias in the average frequency, the 179\,mm coupler introduces a downward drift, and the 149\,mm coupler achieves the smallest $\sigma$, indicating near-ideal symmetric splitting. This optimal coupling length is therefore adopted for all subsequent lattice constructions.

\section{Hofstadter Butterfly Spectrum}

The Hofstadter butterfly spectrum is generated by mapping the realified Hamiltonian~\eqref{eq:quantum Hamiltonian} onto a bilayer lattice of H-shaped resonators and sweeping the gauge field parameter $\theta$ across the interval $[0,\pi]$. For each value of $\theta$, the $2\times 2$ coupling matrices in $H_R$ prescribe the magnitude and sign of every inter- and intra-layer connection, fully determining the lattice geometry. The diagonal blocks of these matrices encode the coupling within each layer, while the off-diagonal blocks govern the inter-layer connections through the bridging channels.

The admissible values of $\theta$ for a lattice with periodicity $L$ (the number of resonators per unit cell along each direction) are $\theta = 2\pi n / L$, where $n$ is a positive integer satisfying $\theta \in [0,\pi]$. Because the Hofstadter spectrum obeys the mirror symmetry $E(\theta) = E(\pi - \theta)$, values beyond $\theta = \pi$ are redundant. Increasing $L$ yields a finer sampling of the $\theta$ axis and reveals progressively more intricate fractal detail in the band structure.

\begin{figure}[b!]
    \centering
    \def\svgwidth{\textwidth}
    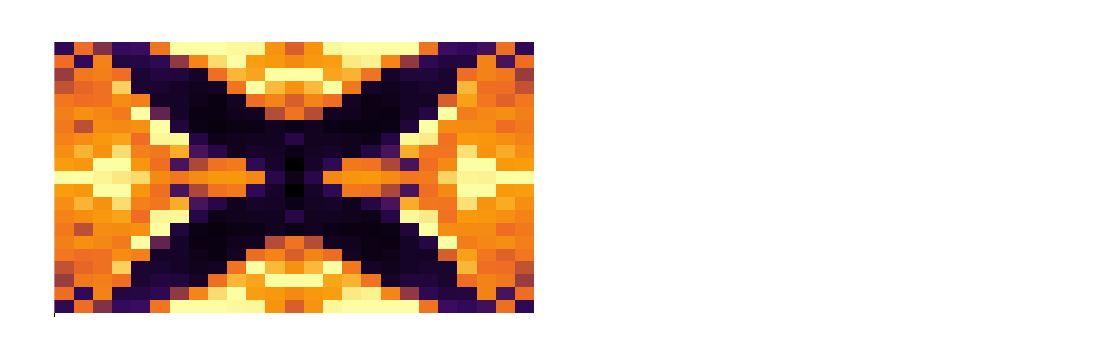
\caption{
\textbf{Hofstadter butterfly spectrum from analytical model and acoustic simulation.}
\textbf{a)} Local density of states (LDOS) computed from the analytical two-layer realified Hofstadter Hamiltonian, $H = \sum e^{i\theta(\mathbf{x} \wedge \mathbf{x}')} |\mathbf{x}\rangle\langle\mathbf{x}'|$, on a 9$\times$9 lattice per layer (162 total sites) with open boundary conditions. The complex Hofstadter Hamiltonian is realified by replacing the imaginary unit with a 2$\times$2 rotation matrix, producing a bilayer system. The x-axis is the gauge field parameter $\theta$. The LDOS is obtained by 2D histogram binning of $(\theta, E)$ pairs with per-$\theta$ column normalization. The spectrum exhibits the expected symmetry $E(\theta) = E(\pi - \theta)$.
\textbf{b)} LDOS from COMSOL finite-element simulation of a 9$\times$9 bilayer coupled acoustic resonator lattice implementing the same synthetic magnetic field. The y-axis shows the physical eigenfrequency range (approximately 2.38--2.46\,kHz). The same binning and normalization procedure as in (a) is used. The agreement between the analytical tight-binding model and the full acoustic simulation confirms that the coupled resonator system faithfully implements the realified Hofstadter model, with the characteristic butterfly band structure and topological gaps clearly reproduced in both panels.
}
    \label{fig:2}
\end{figure}

Figure~\ref{fig:2} presents the Hofstadter butterfly obtained from two complementary approaches. Panel~(a) shows the local density of states (LDOS) computed by analytically diagonalizing $H_R$ on a $9\times 9$ lattice per layer (162 total sites) under open boundary conditions. The LDOS is constructed by 2D histogram binning of $(\theta, E)$ eigenvalue pairs with per-$\theta$ column normalization, producing an intensity map that highlights both the fractal band structure and the states residing within spectral gaps. Panel~(b) displays the corresponding LDOS extracted from full-wave COMSOL eigenfrequency simulations of the $9\times 9$ bilayer acoustic resonator lattice, processed with the identical binning and normalization procedure. The physical eigenfrequencies span approximately 2.38--2.46\,kHz, and the characteristic butterfly pattern---including its self-similar gap structure and the $E(\theta) = E(\pi-\theta)$ symmetry---is clearly reproduced. The quantitative agreement between the two panels confirms that the coupled resonator system faithfully implements the realified Hofstadter model across the full $\theta$ domain.

\section{Spectral and Spatial Characterization at $\theta = 2\pi/3$}

To examine the topological features of the system in detail, we fix the gauge field parameter at $\theta = 2\pi/3$, which corresponds to $n=1$ and $L=3$ in the periodicity relation $\theta = 2\pi n/L$. This choice yields a unit cell of $3\times 3$ resonators per layer---the smallest periodicity that permits three full repetitions of the unit cell within a $9\times 9$ lattice, thereby providing sufficient spatial extent to distinguish boundary-localized states from bulk-extended modes.

\begin{figure}[b!]
    \centering
    \def\svgwidth{\textwidth}
    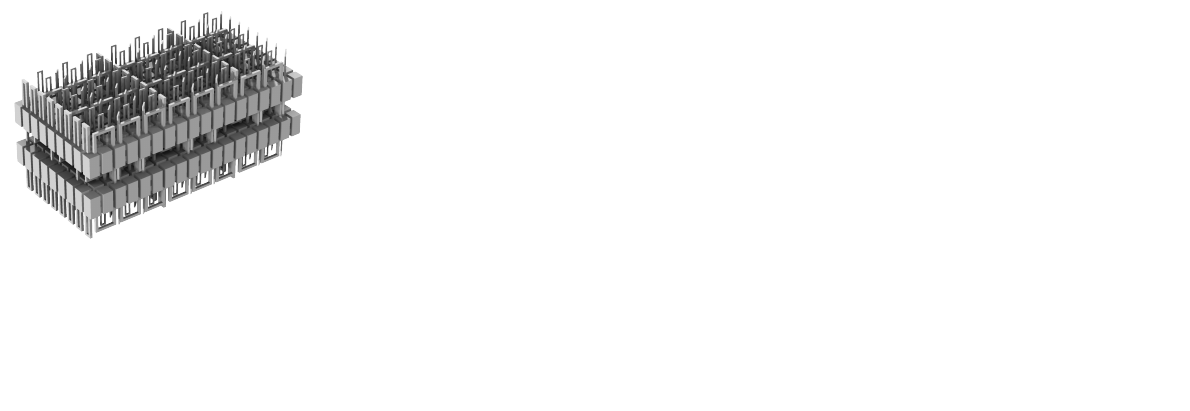
\caption{
\textbf{Spectral and spatial characterization of a 9$\times$9 bilayer acoustic resonator lattice at $\theta = 2\pi/3$.}
\textbf{a)} 3D rendering of the bilayer coupled resonator lattice. Each unit cell is an H-shaped resonator consisting of two square cavities connected by a bridge. The two layers are stacked vertically, with interlayer and intralayer couplings implementing the realified Hofstadter Hamiltonian.
\textbf{b)} Left: COMSOL eigenfrequencies versus solution number for all 162 modes. The spectrum separates into five band regions labeled i--v, with bulk bands (i, iii, v) separated by gap bands (ii, iv). Numbers 1--6 mark individual modes shown in panel (c). Right: density of states (DOS) histogram confirming the band gaps where the DOS drops to zero.
\textbf{c)} Acoustic pressure magnitude $|p|$ for six selected eigenmodes (labeled 1--6) visualized on the 3D lattice structure. Bulk modes (1,3,4, and 6) exhibit pressure distributed across many resonators, while gap modes (2 and 5) show spatially localized profiles.
\textbf{d)} Top-down reconstructions of the 9$\times$9 lattice for each of the five band regions i--v. Each resonator is colored by the summed $|p|$ across all modes within that band, normalized to $[0,1]$. Bulk bands (i, iii, v) show pressure spread across the lattice, while gap bands (ii, iv) exhibit localized pressure distributions characteristic of topological edge states.
}
    \label{fig:3}
\end{figure}

Figure~\ref{fig:3}(a) shows a 3D rendering of the resulting bilayer coupled resonator lattice. Each H-shaped resonator occupies one lattice site, with its two square cavities encoding the two-component internal degree of freedom of $H_R$. The top and bottom layers are stacked vertically, connected by inter-layer couplers whose orientations (same-cavity or cross-cavity) encode the signs of the off-diagonal elements in Eq.~\eqref{eq:quantum Hamiltonian}. Intra-layer couplers link nearest neighbors within each plane according to the diagonal elements.

The eigenfrequency spectrum of this $9\times 9$ bilayer system contains $9\times 9\times 2 = 162$ modes, plotted as a function of solution number in Figure~\ref{fig:3}(b) (left panel). The spectrum organizes into five distinct band regions, labeled i--v: three bulk bands (i, iii, v) that carry a high density of closely spaced eigenfrequencies, separated by two gap bands (ii, iv) in which the density of states drops sharply. The accompanying DOS histogram (Figure~\ref{fig:3}(b), right panel) confirms this structure, with pronounced dips corresponding to the spectral gaps.

Six representative eigenmodes, numbered 1--6 in Figure~\ref{fig:3}(b), are visualized in Figure~\ref{fig:3}(c) through acoustic pressure magnitude $|p|$ maps rendered on the 3D lattice. Modes drawn from the bulk bands (1, 3, 4, and 6) exhibit pressure distributed broadly across the interior of the lattice, consistent with spatially extended Bloch-like states. In contrast, modes 2 and 5, which reside within the gap bands, display pressure profiles that are strongly concentrated along the lattice boundaries while the bulk interior remains largely quiescent. This spatial confinement to the edges is the hallmark of topologically protected edge states.

The distinction between bulk and edge character is further reinforced by the top-down band-region reconstructions in Figure~\ref{fig:3}(d). For each of the five band regions i--v, the acoustic pressure magnitude $|p|$ is summed over all modes belonging to that band and mapped onto a $9\times 9$ grid, with values normalized to $[0,1]$. Bulk bands (i, iii, v) produce pressure that is distributed relatively uniformly across the lattice, whereas gap bands (ii, iv) show pressure concentrated at the boundary sites, forming a clear perimeter pattern. This band-resolved visualization provides compelling evidence that the spectral gaps host edge-localized modes arising from the nontrivial Chern topology of the system.

\subsection{Topological characterization: Chern numbers and bulk--boundary correspondence}
\label{sec:chern}

The edge localization documented in Fig.~3 is suggestive of topological
order, but a quantitative assignment of topological invariants is needed
to establish it. We therefore compute the Chern numbers of the bands and
gaps at the representative flux $\theta = 2\pi/3$ and verify their
consistency with the observed boundary modes through the
bulk--boundary correspondence.

In the gauge of Eq.~(1), the phase accumulated around an elementary
plaquette is $2\theta$, so $\theta = 2\pi/3$ corresponds to a magnetic
flux $\phi = 2\theta/2\pi = 2/3 \equiv -1/3 \pmod 1$ per plaquette, i.e.
a rational flux $p/q$ with $q = 3$. The complex Hofstadter
Hamiltonian~\cite{hofstadter1976} at this flux possesses three magnetic
bands separated by two spectral gaps---precisely the band structure
(regions i, iii, v separated by gaps ii, iv) resolved in the acoustic
eigenspectrum of Fig.~3(b).

We evaluate the band Chern numbers using the gauge-invariant lattice
prescription of Fukui, Hatsugai, and Suzuki, computing the Berry
curvature from link variables on a discretized magnetic Brillouin zone.
The three bands carry
\begin{equation}
  C = (+1,\,-2,\,+1), \qquad \sum_{b} C_b = 0,
  \label{eq:bandchern}
\end{equation}
the canonical result for $|\phi| = 1/3$. The corresponding gap (Hall)
Chern numbers, obtained as the cumulative sum of band Chern numbers below
each gap, are $C_{\rm gap} = +1$ for the lower gap (ii) and
$C_{\rm gap} = -1$ for the upper gap (iv). These values satisfy the
Diophantine relation $j = nq + C_{\rm gap}\,p$~\cite{thouless1982}
($j=1\!:\,1 = 0\cdot 3 + 1$; $j=2\!:\,2 = 1\cdot 3 - 1$), confirming the
assignment.

Crucially, the realified construction of Eq.~(4) does not reproduce a
bare Class~A Chern insulator. Because $H_R$ is real and symmetric it is
invariant under an antiunitary operation, and the onsite generator
$\bigl(\begin{smallmatrix}0&1\\-1&0\end{smallmatrix}\bigr)\otimes\mathbb{1}$
splits it into two decoupled sectors~\cite{barlasProdan2018}. We have
verified numerically that, in the eigenbasis of this generator, $H_R$ is
unitarily equivalent to $H(\theta)\oplus H(-\theta)$ to machine
precision: one sector ($S_+$) emulates the Hofstadter model at flux
$+\theta$ and the other ($S_-$) at $-\theta$. The two sectors therefore
carry opposite Chern numbers,
\begin{equation}
  S_+\!:\ C = (+1,-2,+1), \qquad
  S_-\!:\ C = (-1,+2,-1),
\end{equation}
so that the \emph{total} Chern number of the physical metamaterial
vanishes in every gap, $C_{\rm tot} = C_+ + C_- = 0$, as required by the
antiunitary symmetry. What remains nontrivial is the spin (sector) Chern
number $C_s = (C_+ - C_-)/2 = \pm 1$, which takes the values $-1$ and
$+1$ in gaps ii and iv, respectively. The acoustic system is thus a
realified analogue of a Class~A Chern insulator---a spin-Chern
insulator---whose nonzero $C_s$ protects a \emph{helical} pair of
counter-propagating edge channels in each gap rather than a single
chiral channel.

\begin{figure}[t!]
  \centering
  \includegraphics[width=\columnwidth]{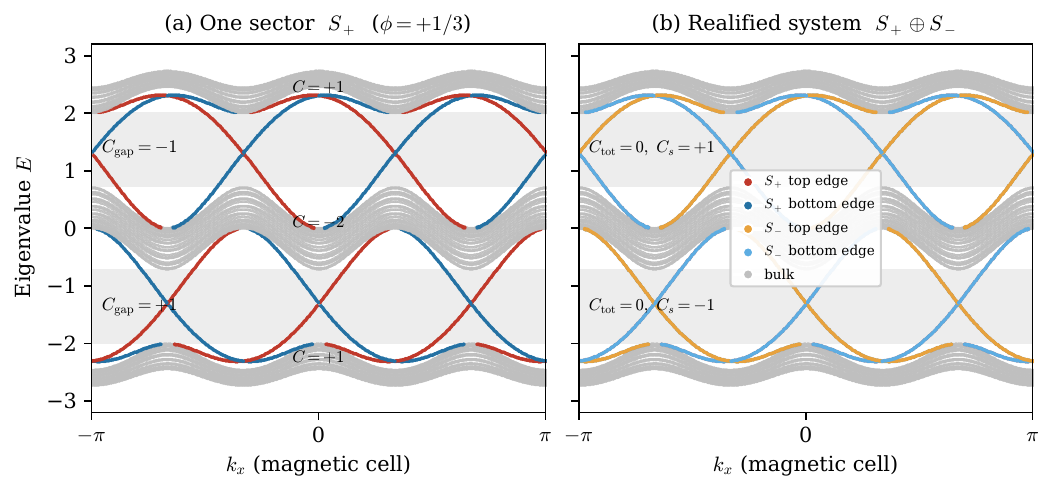}
  \caption{\textbf{Bulk--boundary correspondence of the realified
  Hofstadter lattice at $\theta = 2\pi/3$ ($|\phi|=1/3$).} Ribbon
  dispersion (periodic along $x$ with a $q=3$ magnetic cell, open along
  $y$; bulk states in grey, gap regions shaded). \textbf{(a)} A single
  sector $S_+$: one chiral edge branch threads each bulk gap, localized
  on the top (red) or bottom (blue) boundary with opposite group
  velocity, consistent with the gap Chern numbers $C_{\rm gap}=+1$
  (lower) and $-1$ (upper) and the band Chern numbers $(+1,-2,+1)$.
  \textbf{(b)} The realified system $S_+\!\oplus\!S_-$: the second sector
  ($S_-$, orange/light blue) contributes oppositely dispersing branches
  on each boundary, forming a counter-propagating helical pair in every
  gap. The total Chern number vanishes ($C_{\rm tot}=0$) while the spin
  Chern number $C_s=\pm1$ remains nontrivial.}
  \label{fig:chern}
\end{figure}

This picture is confirmed directly by the bulk--boundary
correspondence~\cite{hatsugai1993,prodanSB2016}. Figure~\ref{fig:chern}(a)
shows the dispersion of a single sector on a ribbon geometry (periodic
along $x$, open along $y$): each gap is traversed by exactly one chiral
edge branch per boundary, with opposite group velocities on opposite
edges, consistent with $|C_{\rm gap}| = 1$. In the full realified system,
Fig.~\ref{fig:chern}(b), the two sectors contribute oppositely dispersing
branches localized on the same boundary, forming the helical pair
guaranteed by $|C_s| = 1$. The single edge-localized mode resolved in
each gap band of Fig.~3(c) (modes~2 and~5) is the acoustically observable
member of such a sector pair. The same counter-propagating edge structure
underlies the dimensionally reduced, one-dimensional quasiperiodic
acoustic realization of Hofstadter physics reported
previously~\cite{ni2019}.

This characterization also clarifies the nature of the topological
protection probed in Sec.~\ref{sec:robustness}. The helical edge pair is
protected by the sector decoupling: perturbations that respect the
two-cavity (sector) structure---such as the removal of a complete unit
cell---cannot mix $S_+$ and $S_-$ and therefore leave the edge modes
intact, whereas perturbations that break this structure could in
principle backscatter the counter-propagating partners. The robustness
demonstrated below should be understood in this symmetry-protected sense.

\section{Robustness Against Structural Defects}
 \label{sec:robustness}

A defining property of topologically protected edge states is their resilience to perturbations that do not close the bulk gap. To test this in our system, we introduce a structural defect into the $9\times 9$ bilayer lattice at $\theta = 2\pi/3$ by removing one complete unit cell (a $3\times 3$ block of resonators) from each layer. All remaining geometric and coupling parameters are held fixed. This controlled perturbation breaks the translational periodicity of the lattice and creates an interior void that the edge modes must navigate around.

\begin{figure}
    \centering
    \def\svgwidth{\textwidth}
    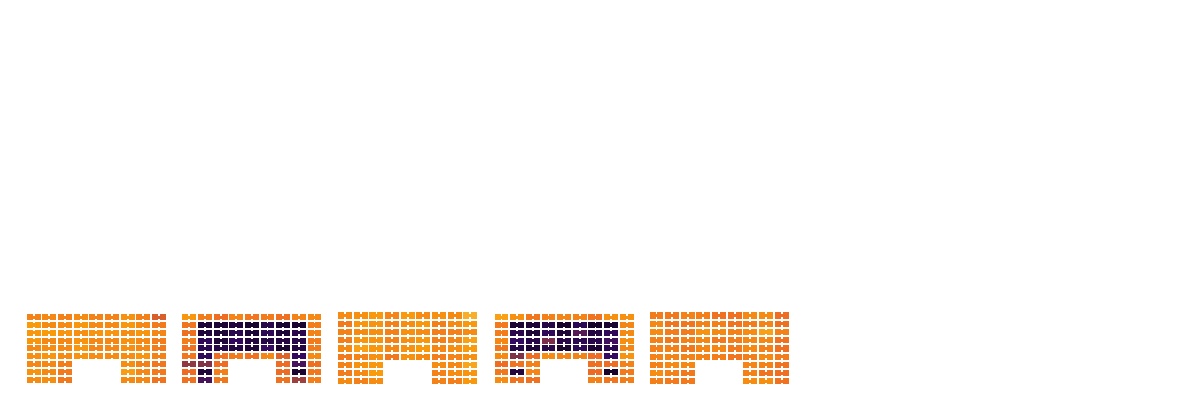
\caption{
\textbf{Defect robustness of the 9$\times$9 bilayer acoustic resonator lattice at $\theta = 2\pi/3$.}
\textbf{a)} 3D rendering of the bilayer lattice with one complete unit cell ($3\times 3$ resonators) removed from each layer.
\textbf{b)} Left: COMSOL eigenfrequencies versus solution number for all 144 modes. The five-band structure (i--v) is preserved despite the defect. Right: DOS histogram.
\textbf{c)} Acoustic pressure magnitude $|p|$ for selected eigenmodes visualized on the 3D lattice. Gap modes remain spatially localized at the boundaries, routing around the void without scattering into the bulk.
\textbf{d)} Top-down reconstructions of the 9$\times$9 lattice for each band region i--v. Gap bands (ii, iv) show edge-localized pressure distributions that bypass the defect site, confirming the topological protection of the edge states.
}
    \label{fig:4}
\end{figure}

Figure~\ref{fig:4}(a) shows a 3D rendering of the defective bilayer lattice, with the void clearly visible. The removal of one unit cell per layer reduces the total mode count from 162 to $(81-9)\times 2 = 144$. Figure~\ref{fig:4}(b) presents the eigenfrequency spectrum plotted against solution number. Despite the missing resonators, the five-band structure (i--v) observed in the pristine lattice (Figure~\ref{fig:3}(b)) is preserved, with bulk bands and spectral gaps remaining at comparable frequency positions. The gap bands continue to host isolated eigenfrequencies consistent with edge-localized states.

Representative pressure magnitude maps for modes drawn from each band region are shown in Figure~\ref{fig:4}(c). Bulk modes again exhibit pressure spread across the lattice interior, while gap modes remain confined to the boundaries. Crucially, the edge states route around the void without scattering into the bulk, maintaining their spatial confinement along the lattice perimeter. The top-down band-region reconstructions in Figure~\ref{fig:4}(d) confirm this behavior: the gap bands (ii, iv) display edge-concentrated pressure patterns that smoothly bypass the defect site, while the bulk bands (i, iii, v) show distributed pressure with the void appearing as a region of zero amplitude.

The persistence of edge-localized modes in the presence of a substantial geometric disruption---the complete removal of nine resonators per layer---provides strong computational evidence for the topological origin of these states. Their robustness is consistent with the bulk-boundary correspondence guaranteed by the nonzero Chern numbers of the underlying Hofstadter Hamiltonian \cite{prodanSB2016, hatsugai1993}.

\section{Experimental Validation}\label{single}

\begin{figure}[t!]
\center{
\begin{tabular}{cccc}
\includegraphics[width=.225\columnwidth]{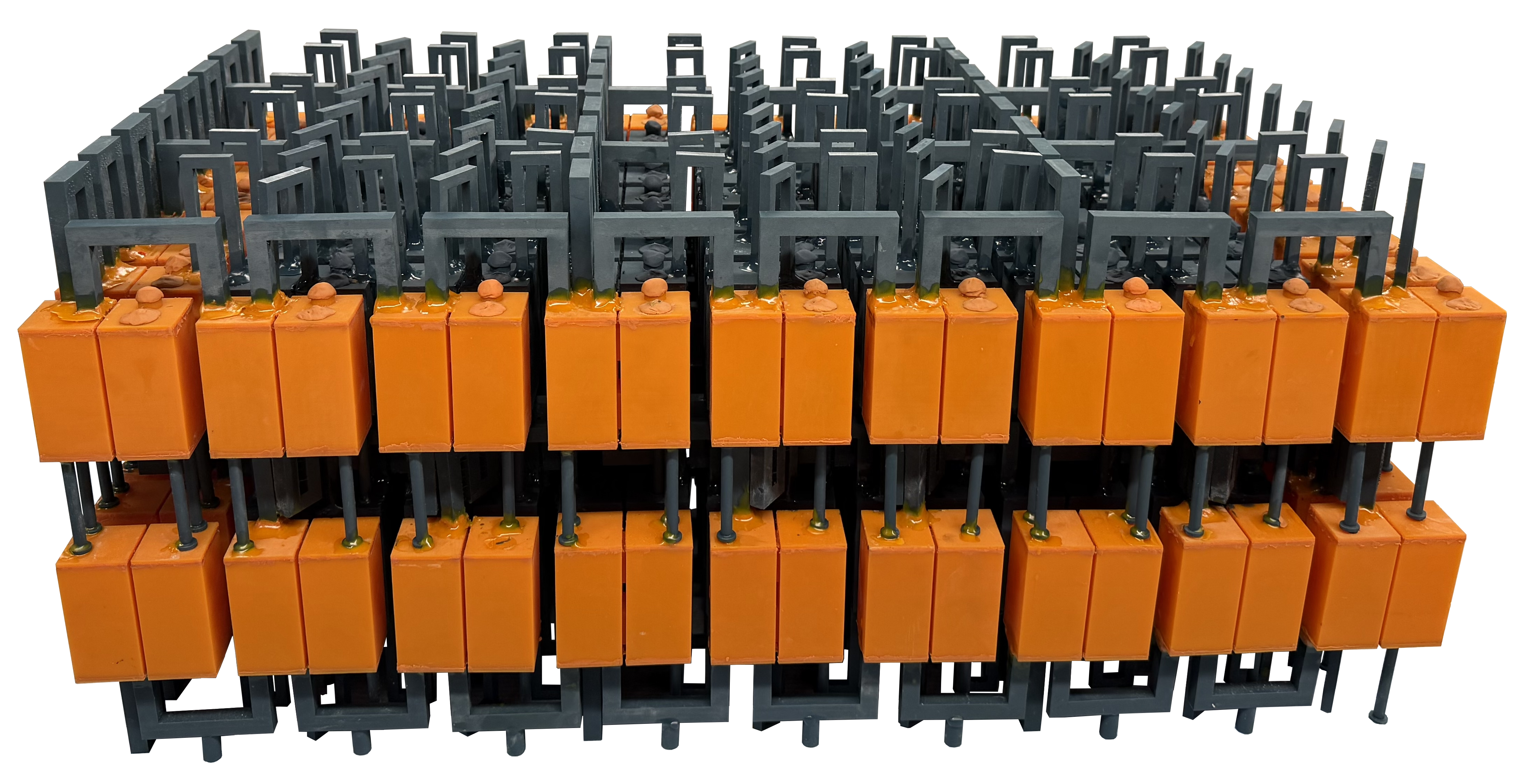} &
\includegraphics[width=.225\columnwidth]{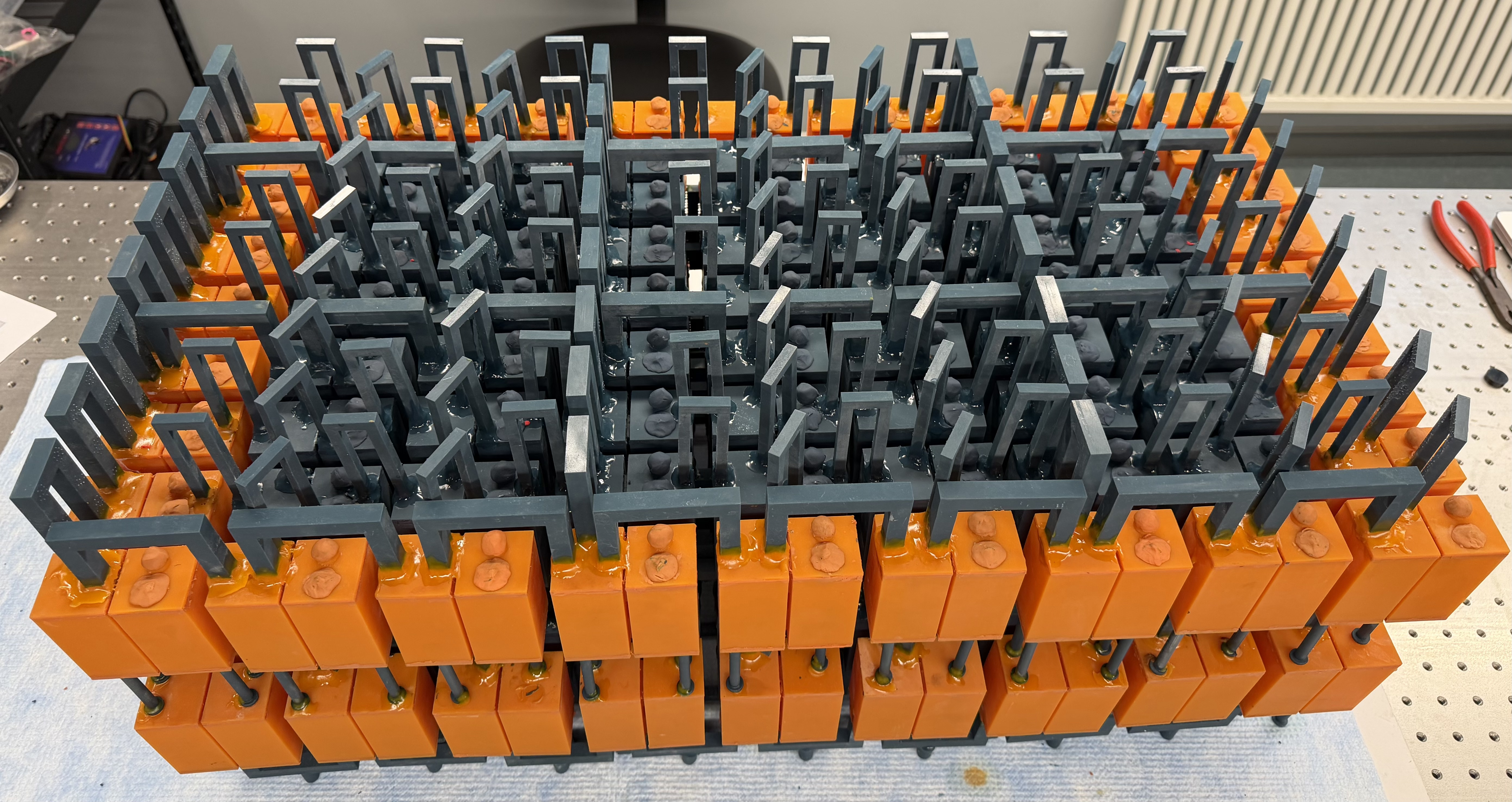} &
\includegraphics[width=.225\columnwidth]{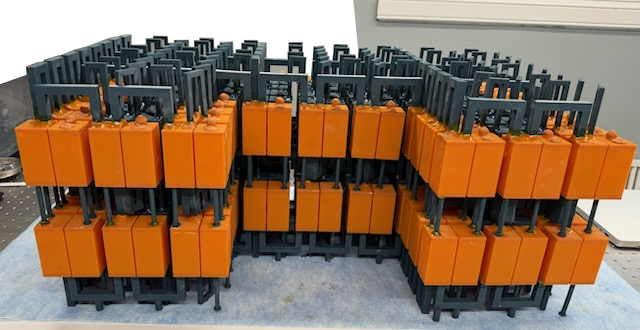} &
\includegraphics[width=.225\columnwidth]{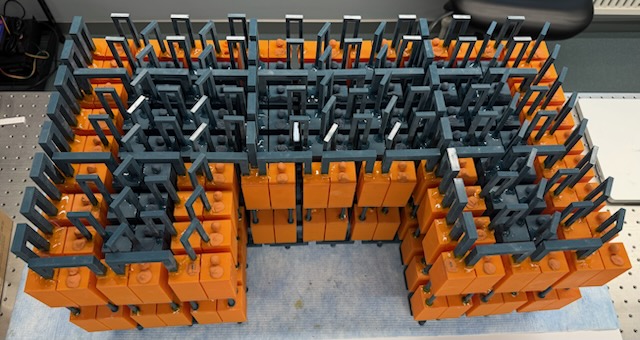} \\
\hspace{8pt} (a) & \hspace{8pt} (b) & \hspace{8pt} (c) & \hspace{8pt} (d)\\\\
 \end{tabular}
}
\caption{\textbf {Experimental model of (a)-(b) the whole $9\times9$ bilayer acoustic resonator lattice and (c)-(d) the defected lattice at $\theta = 2\pi/3$}}
\label{fig:6}
\end{figure}

To confirm the experimental feasibility of the system under study, a model with a defect, shown in Figures~\ref{fig:3}(a)~and~\ref{fig:4}(a), was recreated using 3D modeling and 3D printing of all necessary components. Photographs of the experimental system are shown in Figure~\ref{fig:6}. 

\begin{figure}[t!]
\center{
\includegraphics[width=.9\columnwidth]{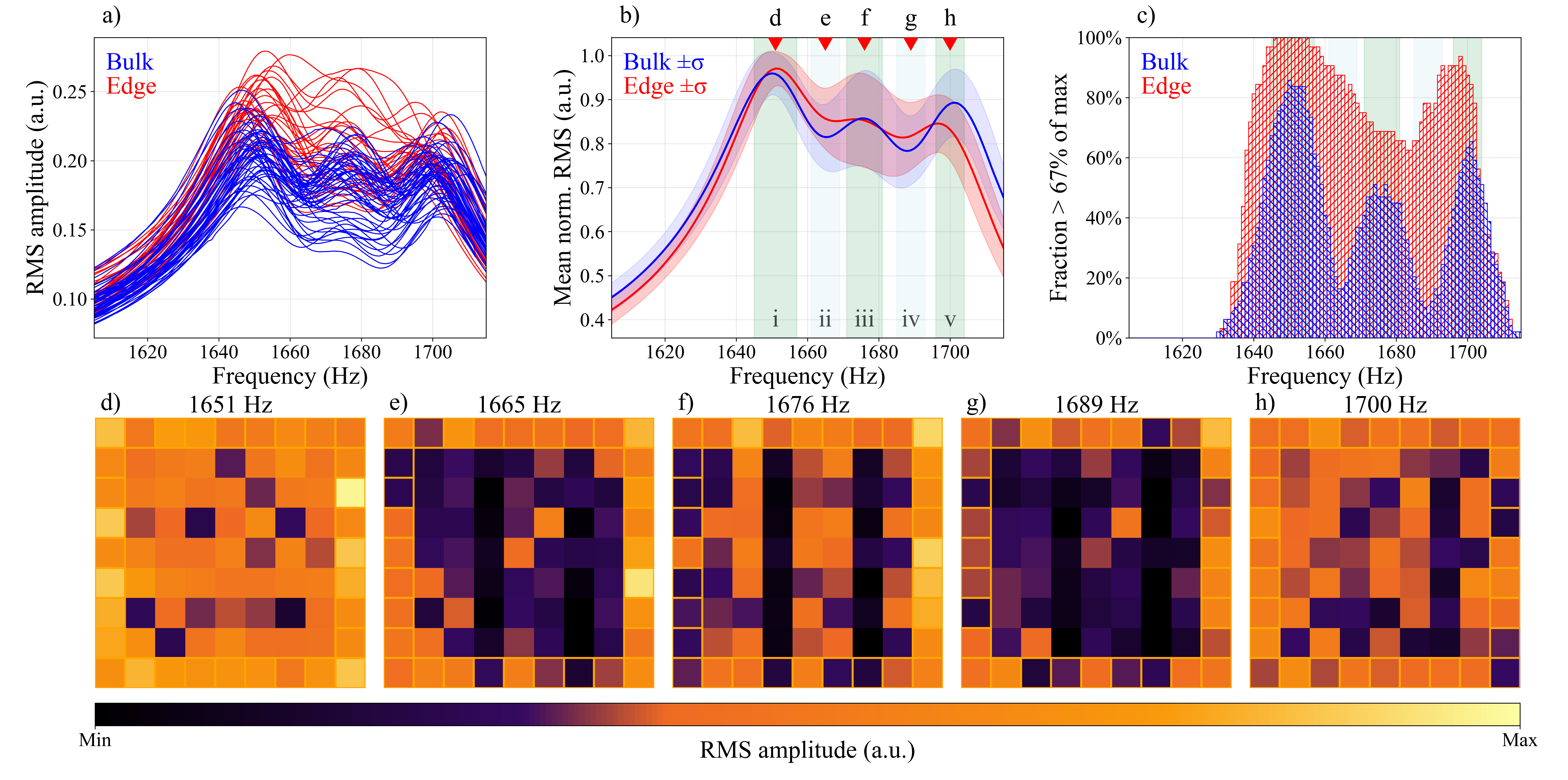}
}
\caption{\textbf{Experimental evidence of existence of the topological five-band region in the $9 \times 9$ bilayer acoustic resonator lattice at $\theta = 2\pi/3$} \textbf{a)} RMS amplitude responses of bulk and edge elements as a function of driving frequency. \textbf{b)} Mean normalized RMS amplitude with confidence intervals. The dependence is divided into five band regions labeled i-v, with bulk bands (i, iii, v) separated by gap bands (ii, iv). Letters d-h mark individual modes shown in corresponding panels below. \textbf{c)} Fractions of edge and bulk resonators whose RMS amplitude at a given frequency exceeds $2/3$ of the global maximum across the spectra of all elements. Blue curves and bars correspond to bulk H-resonators, red curves and bars to edge H-resonators. \textbf{d)-h)} Spatial distributions of the RMS amplitude responses at selected frequencies within each band region. Panels d), f), and h) correspond to bulk regimes, while panels e) and g) correspond to the gap bands. Orange borders highlight the edge elements of the lattice.}
\label{fig:7}
\end{figure}

During the experimental study of the $9 \times 9$ bilayer acoustic resonator lattice at $\theta = 2\pi/3$, swept-frequency RMS amplitude responses were recorded for one cavity of each H-resonator,  and compiled in Figure~\ref{fig:7}(a). As can be seen from this figure, most of the measured frequency responses for the edge resonators have peaks closer to the center (Fig.~\ref{fig:7}(a), red curves) than the spectra corresponding to the bulk elements (Fig.~\ref{fig:7}(a), blue curves). For a more accurate analysis of the acquired responses and to allow identification of a five-band region, the dependence of the mean normalized RMS amplitude response was plotted for the edge and bulk elements (Figure~\ref{fig:7}(b), red and blue curves, respectively). In the intervals corresponding to the gap bands (Figure~\ref{fig:7}(b), blue intervals designated ii, iv), the mean response of the edge resonators clearly exceeds that of the bulk resonators, whereas in the bulk bands (Figure~\ref{fig:7}(b), green intervals designated i, iii, v) both responses are comparable. To illustrate more directly the protection of the bulk resonators against the propagation of acoustic signals at these frequencies, Figure~\ref{fig:7}(c) shows the fractions of edge and bulk resonators whose amplitude at a given frequency exceeds $2/3$ of the global maximum across the spectra of all elements. There are two frequency regions in which the fraction of resonating bulk cavities drops below about $20\%$, while a substantial fraction of the edge cavities still responds strongly. These regions coincide with the gap bands ii and iv. This analysis made it possible to explicitly distinguish these regimes, even taking into account the losses and disorder that arise in experiments with acoustic resonators. 

Moreover, as expected, the bands can be clearly distinguished not only from the averaged normalized spectra of the resonators, but also directly from the spatial distribution of the signal amplitudes at frequencies lying within these intervals. Figures~\ref{fig:7}(d)–(h) show the distributions of the RMS amplitude response at five frequencies corresponding to the five band regions highlighted in Figure~\ref{fig:7}(b). The RMS distributions at frequencies in the bulk bands (Figures~\ref{fig:7}(d),(f),(h)) show high values of acoustic pressure in both the edge and bulk resonators. In contrast, for the edge band frequencies (Figures~\ref{fig:7}(e),(g)), the maximum acoustic pressure is predominantly observed in edge resonators, while the non-zero RMS amplitude response for a few bulk elements arises due to the losses and disorder inherent in the  experimental setup. High losses in the regions of rows 4 and 7, which are clearly visible in Figures~\ref{fig:7}(e)–(g), are associated with the presence of several strong couplings in these regions, which increase the degree of losses relative to other resonators.

\begin{figure}[t!]
\center{
\includegraphics[width=.9\columnwidth]{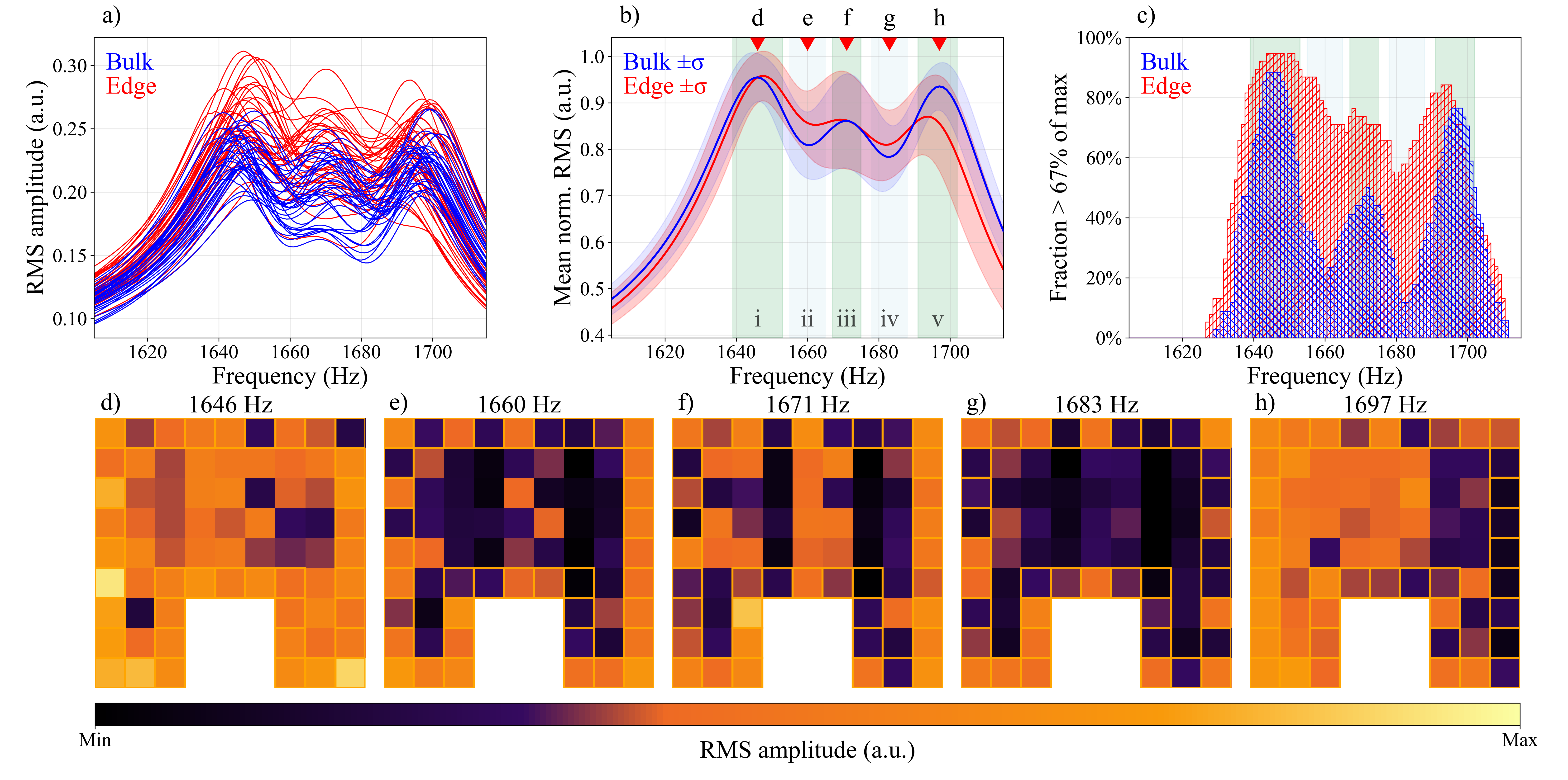}
}
\caption{\textbf{Experimental evidence of existence of the topological five-band region in the $9 \times 9$ bilayer acoustic resonator lattice at $\theta = 2\pi/3$ with the border defect. } \textbf{a)} RMS amplitude responses of bulk and edge elements as a function of driving frequency. \textbf{b)} Mean normalized RMS amplitude with confidence intervals. The dependence is divided into five band regions labeled i-v, with bulk bands (i, iii, v) separated by gap bands (ii, iv). Letters d-h mark individual modes shown in corresponding panels below. \textbf{c)} Fractions of edge and bulk resonators whose RMS amplitude at a given frequency exceeds $2/3$ of the global maximum across the spectra of all elements. Blue curves and bars correspond to bulk H-resonators, red curves and bars to edge H-resonators. \textbf{d)-h)} Spatial distributions of the RMS amplitude responses at selected frequencies within each band region. Panels d), f), and h) correspond to bulk regimes, while panels e) and g) correspond to the gap bands. Orange borders highlight the edge elements of the lattice.}
\label{fig:8}
\end{figure}

The study of the bilayer acoustic resonator lattice with the border defect also yielded results that are in good agreement with those predicted theoretically and obtained in the numerical experiment (see Figure~\ref{fig:4}). The corresponding spectra, spatial distributions of the RMS amplitude responses at selected frequencies within each band region, and histogram are presented in Figure~\ref{fig:8}. 
As for the full system, five bands can be clearly distinguished in the mean normalized RMS spectra (Figure~\ref{fig:8}(b)) by comparing the responses of the edge and bulk resonators: three bulk bands (i, iii, v) and two gap bands (ii, iv).
The histogram in Figure~\ref{fig:8}(c) again shows two regions in which the fraction of resonating bulk cavities is below about $20\%$, coinciding with the gap bands ii and iv. 
The spatial distributions of the RMS amplitude responses at selected frequencies within these five bands confirm leakage into the bulk for i, iii, and v frequency intervals (Figures~\ref{fig:8}(d),(f),(h)) and topological protection of the bulk resonators at ii and iv intervals (Figures~\ref{fig:8}(e),(g)). 
It should be noted that a slight downward shift in frequencies, along with an increase in RMS amplitude, is observed compared with the full system. This effect is attributed not only to the reconfiguration of the system and the possible influence on the remaining elements, but also to the use of a different speaker, which delivered slightly different output values when driven by a signal of the same frequency and voltage as in the case of the full system.

\section{Conclusion}

We have designed and computationally characterized a two-dimensional acoustic metamaterial that realizes Class A topological insulating behavior through a bilayer lattice of double-bridge H-shaped resonators. The coupling architecture implements a realified Hofstadter Hamiltonian \cite{barlasProdan2018}, whose $2\times 2$ matrix structure maps naturally onto the two-cavity geometry of each resonator. By sweeping the synthetic magnetic flux parameter $\theta$ across its fundamental domain, we reproduced the Hofstadter butterfly spectrum---a fractal energy landscape characteristic of 2D lattices in a perpendicular magnetic field \cite{hofstadter1976}---in both the analytical tight-binding model and in full-wave COMSOL finite element simulations. The quantitative agreement between the two approaches validates the use of passive acoustic resonator networks as a faithful platform for emulating Hofstadter physics.

At the representative gauge field value $\theta = 2\pi/3$, the eigenfrequency spectrum of a $9\times 9$ bilayer lattice separates into well-defined bulk and gap bands. The gap bands host spatially localized edge modes that are confined to the lattice boundaries, as confirmed by pressure magnitude visualizations and band-resolved top-down reconstructions. These edge states persist when a complete unit cell is excised from each layer, demonstrating robustness against substantial structural defects consistent with the topological protection guaranteed by nonzero Chern numbers. Moreover, the configuration was tested within the experiment using the fabricated acoustic lattice sample for the full system and for the system with the border defect. The sample was created using a 3D printer, and the transmission spectra were measured for all H-resonators of the top layer. The analysis of the spectra revealed the five-band structure, which include the bulk bands and the gap bands exhibiting localized pressure distributions characteristic of topological edge states.

Our results establish a computational framework for designing acoustic metamaterials with synthetic gauge fields and Class A topology, complementing recent experimental advances in Hofstadter physics across electronic \cite{nuckolls2025}, phononic \cite{daviesMorini2026}, and optomechanical \cite{slim2025} platforms. The double-bridge H-resonator geometry introduced here offers a versatile building block for future implementations targeting other topological phases within the classification table, and for scaling toward larger lattices where richer fractal structure and higher-order topological phenomena may emerge.

\section{Materials and Methods}

\subsection{Numerical Simulations}

All acoustic simulations were performed using the Pressure Acoustics, Frequency Domain interface of COMSOL Multiphysics version 6.3 \cite{comsol}, with the built-in Eigenfrequency Solver used to extract the resonant modes of each lattice configuration. The lattice geometry was constructed programmatically through the LiveLink\textsuperscript{TM} for MATLAB interface \cite{MATLAB}, which enabled a systematic algorithm for generating the full bilayer structure as a function of the gauge field parameter $\theta$ and the lattice size $L$.

The geometry generation algorithm operates in two stages. First, the connectivity between all resonator sites in the bilayer lattice is extracted from the realified Hamiltonian~\eqref{eq:quantum Hamiltonian} for a given $\theta$ and $L$. This connectivity map specifies which resonators are coupled, whether the coupling is intra-layer or inter-layer, and what sign (positive or negative) each hopping element carries. Second, the physical 3D geometry is assembled in COMSOL by loading a template H-resonator model via \texttt{mphload}, then programmatically placing resonators and creating coupling tubes according to the connectivity map. The top layer is generated by mirroring the bottom layer, as both share identical connectivity. The completed model is saved with \texttt{mphsave} and passed to the solver.

Air was assigned as the acoustic medium throughout all domains (resonators and couplers) using the COMSOL materials library. Sound hard boundary conditions (zero normal velocity) were imposed on all exterior surfaces. A free tetrahedral mesh was employed, with refinement tests confirming that an average element quality of approximately 0.67 provides adequate numerical accuracy and stability. The number of requested eigenfrequencies equals the total number of resonators in the bilayer system; for example, a $9\times 9$ bilayer yields 162 eigenfrequencies. No perfectly matched layers were included, as the eigenfrequency analysis does not involve radiating or absorbing boundary conditions.

For configurations requiring periodic boundary conditions (PBCs), additional geometric modifications were implemented. Coupling tubes extending beyond the periodic boundaries were bisected using work planes, and the resulting faces were designated as Floquet periodic boundary pairs with the wavevector set to zero. Conformal meshes on opposing periodic faces were ensured by meshing one face first and copying it to the opposite face. This procedure was applied independently along the $\hat{x}$ and $\hat{y}$ directions.

\subsection{Experiment}

The acoustic resonators and couplers were fabricated using a FormLabs Form 4L stereolithography (SLA) 3D printer which offers an XY resolution of $46$ $\mu m$ and a layer thickness (Z resolution) of $100$ $\mu m$. Two distinct photopolymer resins supplied by FormLabs were employed to facilitate visual differentiation between resonator types: blue resin was used for the bulk H-resonators, while orange resin was used for the edge H-resonators. It should be noted that the different colors of the resins, and consequently their slightly different compositions, has no effect on the acoustic properties of the resonators and thus do not introduce additional inhomogeneities into the experimental system. The wall thickness of the individual resonators was chosen to be $2~mm$, ensuring sufficient structural rigidity while minimizing unwanted acoustic absorption. The couplers were designed with a wall thickness of $1~mm$ to facilitate controlled energy transfer between the resonators.

To enable acoustic measurements, small apertures were integrated into one cavity cap of each H-resonator, allowing insertion of compact omnidirectional condenser microphones (Audix TM1 Plus) and micro-speakers (model SP-1003, Digi-Key) for excitation and response recording. The self-resonance frequency of the microphones is $1350$ Hz, which lies outside the measured frequency range of $1600–1720$ Hz covering the topological five-band region of interest.

Particular attention was paid to the integrity of the coupling channels, as these regions are critical for the topological properties of the system. To mitigate acoustic leakage and minimize losses at these junctions, the joints were carefully sealed with modeling clay and subsequently reinforced with an additional layer of uncured photopolymer resin, which was cured under a high-intensity ultraviolet (UV) lamp for polymerization. Despite these precautions, residual acoustic losses and fabrication imperfections were unavoidable; however, they did not significantly affect the observed topological features.

\section{Acknowledgements}
C.P. and E.R. acknowledges support from the U.S. National Science Foundation through grant CMMI-2414984, and by the U.S. Army Research Office through contract W911NF-25-1-0050. E.P. acknowledges support from the U.S. National Science Foundation through grant CMMI-2131760, and by the U.S. Army Research Office through contract W911NF-23-1-0127. M.R., N.H.P., and C.R. acknowledge support from the National Science Foundation through Award No.\ 2131758.

\section{Author Contributions}
J.S. and E.R. conducted the COMSOL-based computational modeling and data analysis. C.P., E.P., and M.R. conceived the project, provided the theoretical framework, and supervised all aspects of its development. C.R. and N.H.P. contributed to numerical modeling and simulation using both COMSOL and MATLAB environments. E.R. created the experimental model and analyzed the experimental data. All authors contributed to the analysis and interpretation of the results and were involved in reviewing and writing the manuscript.

\section{Competing Interests}
The authors declare no competing interests.

\bibliographystyle{naturemag}
\bibliography{references}

\end{document}